# Capturing the Tax-Revenue Bracketing System via a predator-prey model: Evidence from South Africa

Leonard Mushunje

Columbia University, Department of Statistics

10027, NY, NY, USA

lm3748@columbia.edu

**Abstract**

Revenues obtained from the corporate tax heads play significant roles in any economy as they can be prioritized for producing public goods and employment creations, among others. As such, corporate tax revenue should be paid enough attention. This study, therefore, explores the tax-revenue harvesting system of an economy where we focused on the corporate tax head. The system comprises three players; the government and formal and informal firms. We applied the predator-prey model to model the effect of the government-gazetted tax rate on corporate survivability. It is a new approach to modeling economic system relations and games. Critical combinatory points are derived, with stability analysis provided after that. Dynamics associated with the tax-revenue system are established and critically analyzed. Lastly, we provide the mathematical way the system can be optimized for the government to harvest as much Revenue as possible, including optimal conditions.

1. ## Introduction

All forms of formal and informal businesses are the primary and critical sources of Revenue for the government. This often comes in the form of tax payments. However, the number of sounding firms with better profit centers could be much higher. As such, it could be better, especially for emerging governments like South Africa to have some firms shutting down and some dying out. Of course, the government has some revenue avenues besides tax heads, like profits from parastatals and balance of payments (trade surpluses), but taxes are the primary and fundamental pillars. This gives a clear picture of why better attention should be paid to tax circulation and its harvest in the economy. This study will dwell on the mathematical ways of analyzing tax revenue harvesting in South Africa. It introduces a new way of tax revenue harvesting through the interaction of the government and the firms (formal and informal). We shall employ the predator-prey model of the Lotka-Volterra type and consider its application to our problem. The concept of tax-revenue harvesting is familiar—however, neither is the predator-prey model. But the application of the model in specific tax-revenue harvesting is novel. As in the literature, Chenglin Li (2019), looks at the Spatiotemporal pattern formation of a ratio-dependent invasion-diffusion predator-prey system with disease in the predator case. Their findings state that in a wide space for the predator and the prey to diffuse by the self-pressure and with a little tendency of predator to catch prey, they can't coexist. There isn't endemic disease extensively, but they can coexist. The endemic disease exists broadly when the coefficient of invasion-diffusion for predators is big enough, with other random diffusion coefficients being fixed and satisfying certain conditions. In addition, Lee, Lee & Oh, (2005) have used predator-prey models to analyze many economic situations, including the Korean stock market and Modis, (2003) studies the competition between ballpoint and fountain pens. More profoundly, a fundamental model in economics is the Goodwin model. The model was developed to model the economic fluctuations regarding real wages and actual employment. Mathematically, the Goodwin model can easily be related to the Lotka-Volterra model, see Vadasz (2007). Seong-Joon Lee, Deok-Joo Lee, and Hyung-Sik Oh (2005) are direct examples of predator-prey models in the economic field.

Moreover, Gracia (2004) fits the business cycle to the Lotka-Volterra predator-prey model and his results agreed with the efficient markets hypothesis. Additionally, Berestycki and Zilio, (2019) introduce a model to shed light on the emergence of territorial behavior in predators with the formation of packs. In their paper, they consider the situation of predators competing for the same prey (or spatially distributed resource). Their results suggest that solid competition between groups of predators leads to the formation of territories. Not related to economics, we have several extensive studies on the application of the predator-prey models, see (Brown (1964), Wilson (1975). Krebs (1971), Hixon (1980), Johnson and Gaines (1990), Larsen and Boutin (1994), and Dancer and Du (1994). Above all, the reader is recommended to read Lotka-Volterra, (1925 and 1926). The present study aims to apply the predator-prey model to explore and study the tax-revenue bracketing system. There needed to be direct literature on our study.

## 2. Methodology

### 2.1. Data

We used the tax heads' data and corresponding revenue data for South Africa for a period spanning from 1974 to 2021, collected from World Bank statistics. The data can be freely accessed at World Bank-South Africa tax and GDP statistics website. We had a few missing data that we accounted for through the interpolation (KNN) approach; otherwise, the data was sufficient for our analysis.

### 2.2. Predator-Prey model

We constructed our model based on the following assumptions. This is a way of simplifying our model applications and clarifying our application. Thus;

1) At time t, the firms (population) are divided into two (formal and informal businesses).

$$F(t) + \overline{F(t)} = P(t),,,,,,,,,,,(1)$$ where,

F(t) is the total number of formal firms in an economy, $\overline{F(t)}$ is the totality of informal firms in an economy, and $P(t)$ is the sum of the two firm categories.

In literal terms, equation (1) represents a two-sector business economy. This is comprised only of prey components. We then add the third component, which is the predator component. The resulting equation (2) below represents the three-sector business economy.

$$F(t) + \overline{F(t)} + G(t) = P(t),,,,,,,,,,,(2).$$

From the formulated equations, it is essential to note that the firms represent the prey while the government is the predator.

2) We shall define the predator (government) in terms of the tax rate, that is, the corporate tax rate it imposes on the firms and the prey (firms) in terms of their profits.

Both formal and informal firms experience and get affected by the tax rate imposed by the government. As such, their profits and revenues experience some changes at different rates. Thus, the rate of change in earnings of both informal and formal firms as a result of the corporate tax rate at any time t is given by the following general forms respectively;

$$\frac{d\overline{F_p(t)}}{dt} = r\overline{F_p(t)}\left(1 - \frac{\overline{F_p(t)} + \gamma F_p(t)}{K}\right), \ldots\ldots\ldots\ldots\ldots\ldots (3) \text{ and}$$

$$\frac{dF_p(t)}{dt} = rF_p(t)\left(1 - \frac{F_p(t) + \gamma \overline{F_p(t)}}{K}\right), \ldots\ldots\ldots\ldots\ldots\ldots (4)$$

Where γ is the effect of the corporate tax rate on the internal firm's profit accumulations, $K$ is the maximum possible profit margins for the firms, $r$ is the intrinsic profit growth rate, and t is time.

The government harvests its tax revenue from formal and informal company profits, with the informal firms being its mainstream. The reason is their potential to earn high-profit margins than the formal ones, and the government requires less handling time to harvest from the formal ones as they are already the key shareholders and operators. Whenever the government imposes a tax rate, the companies react. This mainly arises from the rationality principle and the utility theory-self-interest by Adam Smith. It is, therefore, essential to analyze all the possible reactions of the prey (firms) following the pre-taken action of the predator (government). We shall examine and determine the reactions using the Holling type 1 and 2 functional response functions. Type 1 will be used in specific legal firms, while type 2 will be for informal ones. The functions are defined as follows:

$$\begin{cases} \tau F_p \ldots\ldots\ldots\ldots\ldots Holling\ type\ 1 \\ \frac{\alpha \overline{F_p}}{\rho + \overline{F_p}} \ldots\ldots\ldots Holling\ type\ 2 \end{cases} \ldots\ldots (5)$$

Here, τ and α are the maximum tax-revenue harvesting/capturing rates for formal and informal firms respectively.

ρ is half of the saturation constant that is $\rho = \frac{K}{2}$

3) We further assume that the shutdown rate of the heavily taxed firms is $\sigma$. Therefore, the rate of change of the formalized and non-formalized firms can be expressed separately as;

$$\frac{d\overline{F_p}(t)}{dt} = r\overline{F_p(t)}\left(1 - \frac{\overline{F_p(t)} + \pi F_p(t)}{K}\right) - \alpha\overline{F_p(t)}F_p - \frac{\beta \overline{F_p}G(t)}{a + \overline{F_p}} \quad \ldots (6)$$

$$\frac{dF_p}{dt} = \alpha\overline{F_p(t)}F_p - \gamma F_p G(t) - \sigma F_p \ldots\ldots\ldots\ldots\ldots, . (7)$$

In all cases, the government has some alternative revenue. The revenue growth due to these sources is denoted by $d_r$ and further letting $l$ and $m$ be the respective contributions of government revenue from informal and formal firms. Additionally, a negative impact $n\gamma I (n \geq 0)$ on the government revenue totals due to the shutdown and slowdown of some informal and formal firms is assumed. The final derived model for the government (predator) revenue structure is presented as:

$$\frac{dG_r}{dt} = \frac{l\beta \overline{F_p}G(t)}{a + \overline{F_p}} + (m - n)\gamma F_p G_r(t) + d_r\left(1 - \frac{\overline{F_p(t)} + \pi F_p(t)}{K}\right)G_r(t) - \mu G_r(t) - \delta G_r^2 \ldots (8)$$

Where μ is the decay rate of the government's tax-revenue capacity and σ its decay density

Now, combining all the constructed differential equations, assumptions, and conditions, we come up with the final revenue-economic system as below:

$$\begin{cases} \frac{d\overline{F_p}(t)}{dt} = r\overline{F_p(t)}\left(1 - \frac{\overline{F_p(t)} + \pi F_p(t)}{K}\right) - \alpha\overline{F_p(t)}F_p - \frac{\beta \overline{F_p}G(t)}{a + \overline{F_p}} \\ \frac{dF_p}{dt} = \alpha\overline{F_p(t)}F_p - \gamma F_p G(t) - \sigma F_p \\ \frac{dG_r}{dt} == \frac{l\beta \overline{F_p}G(t)}{a + \overline{F_p}} + (m - n)\gamma F_p G_r(t) + d_r\left(1 - \frac{\overline{F_p(t)} + \pi F_p(t)}{K}\right)G_r(t) - \mu G_r(t) - \delta G_r^2 \end{cases} \ldots (9)$$

Subject to: $\overline{F_p(0)}, F_p(0), and\ G_r(0)$, which are more significant than zero, are all positive numeral values at the time, $t = 0$. We note that equations 6 to 8, and the combined system of equations (system 9) are positively bounded. Thus, initializing the modelling system/process with significantly positive values guarantees positively bounded solutions to our ordinary differential equations. Again, note that, $\tau$ and $\alpha$ are the maximum tax-revenue harvesting/capturing rates for formal and informal firms respectively.

## 3. Results and Findings

### 3.1. Empirical analysis

We provide an analysis using the Tax head's data and its contribution to Revenue output in South Africa. Corporate taxes, income taxes, excise, and value-added tax are considered in this section.

Figure 1: Tax head Composition

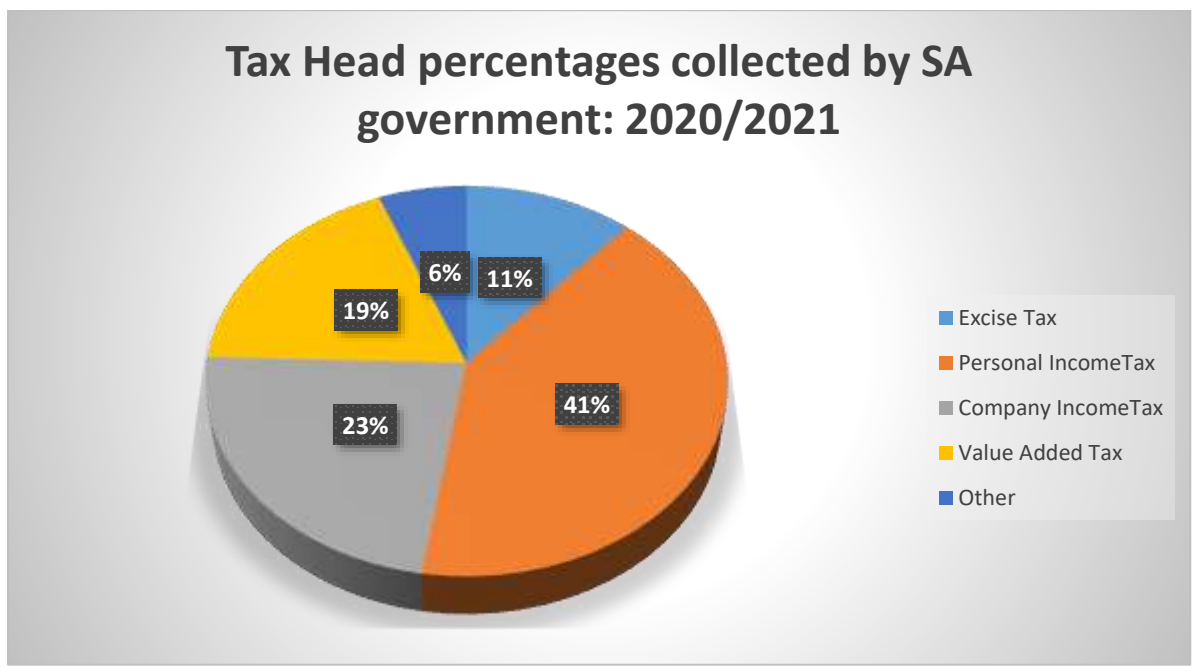

From the pie above, Personal Income tax has the highest proportion, indicating that individual income is more highly taxed than other entities, with an annual rate of 41%. In principle, we expect to see high revenue streams from this head than the rest. Company taxes are the next highest with a rate of 23%, and the other avenues (capital gains taxes and so on) are the least counted. On the other hand, VAT is relatively moderate, together with the excise taxes, with weights of 19% and 11%, respectively. In our study, it is clear that company taxes play a vital role in the revenue harvests of the governments. Below we then provide the overall contribution of the tax heads to revenue streams, quoted in GDP terms.

Figure 2: Tax head contributions to Total Revenue

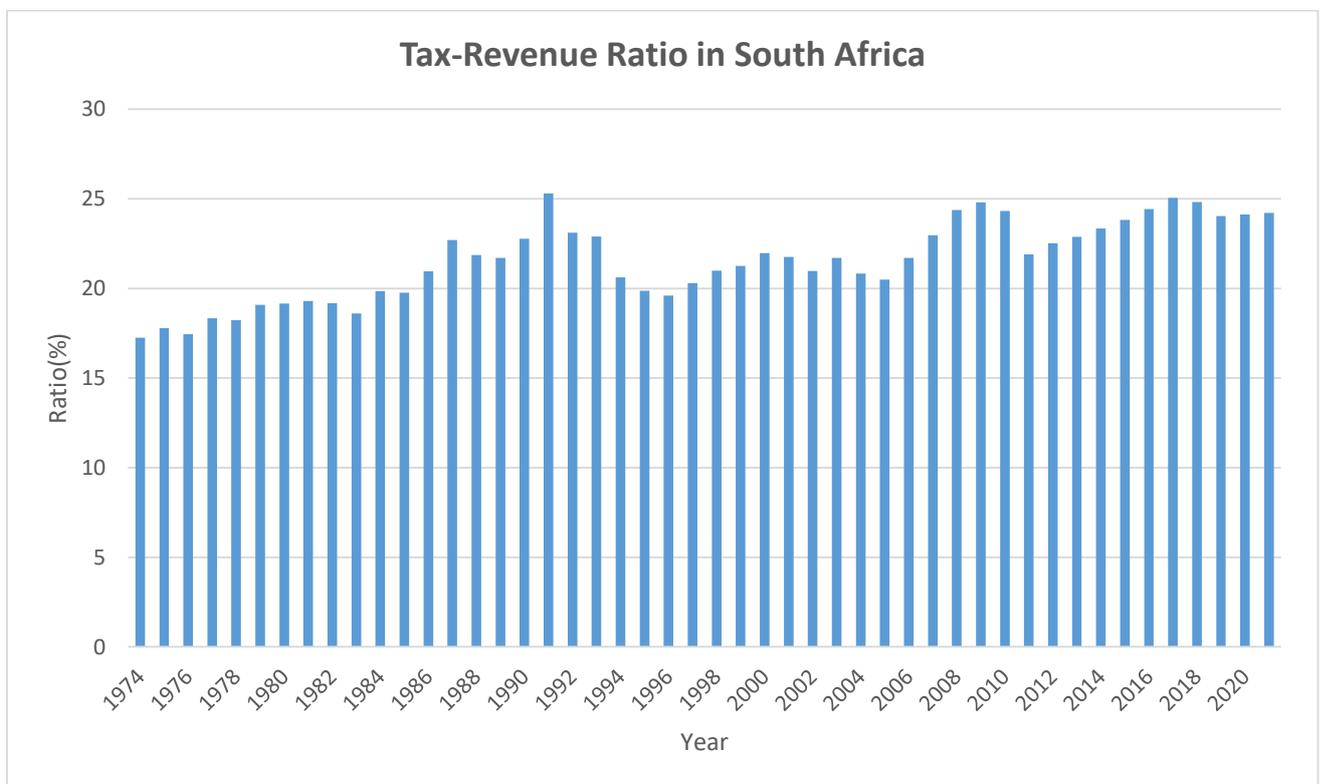

The chart shows the contribution of tax heads to the South African revenue streams. We computed and charted the tax-revenue ratios starting from 1974 to 2021. The balance has been relatively constant over the entire period under study, with the highest percentage picking up to 25% in 1991. Generally, tax revenue has been highly recorded in South Africa and is considered one of the main revenue avenues even in other countries like Zimbabwe.

Figure 3: Company Tax and Income Tax Comparison

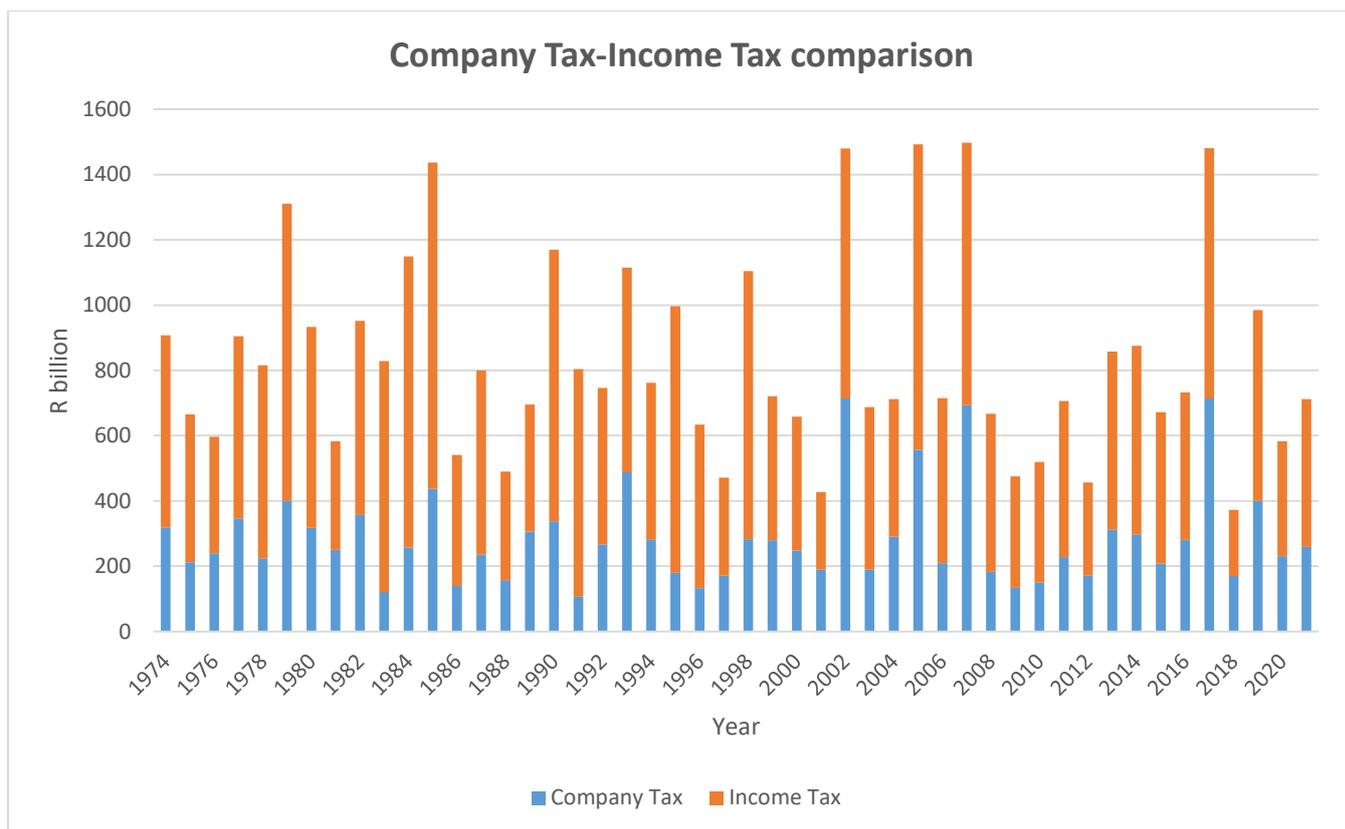

The chart above compares the two primary tax revenue avenues, income and company taxes, from 1974 to 2021. Income tax has always been higher than corporate tax, a phenomenon that holds for most African economies. This is consistent with the above results: the personal income tax revenue ratio is comparatively higher than other tax heads. However, the differences are more minor than in some other cases, indicating that corporate taxes are, in some ways, the most attractable revenue sources to the government and the country at large. Below we present the analysis of the model shown earlier.

### 3.2. Dynamical behavior of the tax-revenue harvesting system

This section aims to make a mathematically based trend/change analysis of the tax-revenue bracketing system. We shall look at; system boundedness, system equilibria, and system stability analysis.

### 3.2.1. System boundedness

We explore mathematically whether the tax-revenue system is bounded or not. We proved that the system in (9) above is uniformly bounded. As a preliminary, we shall define a new systemic function as; $X = \bar{F}_p + F_p + \frac{1}{l}G \ldots \ldots (10)$ assuming that $l$ is the government contribution to the economic population (in a 3-sector economy). So we are adding a component to the 2-sector economy. Let's now find the time derivative of (10) along the solution of the tax-revenue bracket system (9). The derivative is given by;

$$\frac{dX}{dt} = r\overline{F_p(t)}\left(1 - \frac{\overline{F_p(t)} + \pi F_p(t)}{K}\right) - \sigma F_p\left(1 - \frac{m-n}{l}\right)\gamma \bar{F}_p G$$
$$+ \frac{d}{l}\left(1 - \frac{\overline{F_p(t)} + \pi F_p(t)}{K}\right)G\bar{F}_p - \frac{\mu}{l}G - \delta lG \ldots \ldots \ldots \ldots \ldots (11)$$

Now, we decompose (11) by considering the minimum value of $G$ as $G = min\{\sigma, \frac{\mu}{l}\}$ and if we consider $l + n < m$ while taking $X = \bar{F}_p + F_p + \frac{1}{m-n}$. Consequently, if we take the time derivative of X, we obtain the following inequality.

$$\frac{dX}{dt} + gX \leq \bar{F}_p\left[r\left(1 - \frac{\bar{F}_p}{K}\right) + g\right] + \frac{1}{l}G(d - \delta G) \ldots \ldots \ldots \ldots \ldots \ldots \ldots \ldots (12)$$

From (12), we can safely find the maximum values concerning the tax-revenue harvesting limits with time under the maximum capacity constraint (K). By function, the maximum values are

$\max\{r\bar{F}_p\left(1 - \frac{\bar{F}_p}{K}\right) + g\bar{F}_p\} = \frac{K(r+g)^2}{4r}$ and $\max\{(dG - \delta G^2)\} = \frac{d^2}{4\delta}$. Thus using

(12), we have $\frac{dX}{dt} + gX \leq \frac{K(r+g)^2}{4r} + \frac{d^2}{4l\delta}$, and if we let $Z = \frac{K(r+g)^2}{4r} + \frac{d^2}{4l\delta}$ be constant, we will finally have the following inequality;

$\frac{dX}{dt} + gX \leq Z$. This inequality is solved using the method of differential calculus involving inequalities as in the theorem of differential inequality, Birk-hoff and Rota (1953). Using theorem 3.1 above, we obtain

$0 < \bar{F}_p(\bar{F}_p, F_p, G) \leq \frac{Z}{g}(1 - e^{-gt}) + \bar{F}_p\left(\bar{F}_p(0), F_p(0), G(0)\right)e^{-gt} \ldots \ldots \ldots (13)$ and provided that $t \to \infty$ we get $0 < X \leq \frac{Z}{g}$. Conclusively, from this, all analysis, the solutions to the tax-revenue system (9) originate in $\{R_3^+/0\}$, which are antipodal points and are confined in the region, $V = \left\{ (\bar{F}_p, F_p, G) \in R_3^+ : X = \frac{Z}{g} + \epsilon \right\} \ldots\ldots\ldots\ldots\ldots\ldots (14)$. Therefore, the system (9) is always uniformly bounded for all $\epsilon > 0$ and $t \to \infty$.

### 3.2.2. System Equilibria and Stability analysis

We analyzed our tax-revenue bracket system to identify some points of equilibrium that we refer to as critical points. We obtained six different equilibrium points, each with a different meaningful interpretation. These are;(i) The trivial equilibrium $E_0(0,0,0)$.(ii) The boundary equilibrium $E_1(K,0,0)$.(iii) The firm-free equilibrium $E_2(0,0,G_0)$, where, $G_0 = \frac{d-\mu}{\delta}$. This equilibrium is feasible if $d > \mu$, i.e., if the growth rate of government revenue due to the alternative sources is greater than that of its corporates (parastatals). This leads to a point: $E_0$ if $d \leq \mu$.

(iv) We also have a defined non-coexistence free point: $E_3(\bar{F}_1, 0, G_1)$, where $\bar{F}_1$ is the positive root extracted from the following equation: $\bar{F}^3 + l_1\bar{F}^2 + l_2\bar{F} + l_3 = 0,,,,,,,,, (15)$, where,

$$l_1 = \frac{2ar\delta - d\beta - Kr\delta}{r\delta}$$

$$l_2 = \frac{ar\delta - a(d\beta + 2Kr\delta) + \beta K(d + l\beta - \mu)}{r\delta}$$

$$l_3 = -\frac{aK(ar\delta + \beta\mu - d\beta)}{r\delta}$$

and $G_1 = \frac{r}{\beta}\left(1 - \frac{\bar{F}_1}{K}\right)(a + \bar{F}_1)$. The positive root of (15) is obtained under the sufficient condition; $a > K + \frac{d\beta}{(r\delta)}$ for $(r\delta) \neq 0$. Since the $\bar{F}_1$ solution to (15) is not defined, we get $E_3$ in the reduced form of $E_2$, conditionally upon

$d > \mu$. (v) On the other hand, the government's corresponding point is given by: $E_4(\bar{F}_2, F_2, 0)$, where $\bar{F}_1 = \frac{\sigma}{\alpha}$ and $F_2 = \frac{r(\alpha K - \sigma)}{\alpha(K\alpha + r\pi)}$ for $\sigma < \alpha K$,. If the condition fails, we obtain the threshold boundary equilibrium point: $E_1(K, 0, 0)$. (vi) Finally, we have the equilibrium point: $E_*(\bar{F}_*, F*, G_*)$. This is non-trivial, and unless only numerical methods are used, one can only obtain one critical value $\alpha_{\max}$ of $\alpha$. Explicit and closed-form solutions are not easy to find, but finding them is fundamentally equal.

### 3.3.3. Stability analysis of equilibrium points

Local stability

The local stability analysis of our obtained critical points is best and well explained in theorem 3.1, stated below;

**Theorem 3.3.1**

At different equilibria, the system (9) has the following behavior;

(i) An unstable trivial point $E_0(0,0,0)$

(ii) A locally asymptotic bounded threshold point: $E_1(K, 0, 0)$. This holds under the condition: $\sigma > K\alpha$ and $\mu > \frac{K\beta}{(a + K)}$.

(iii) A locally asymptotic stable point $E_2(0, 0, G_0)$ under the following condition:

$d > \mu + \frac{ar\delta}{\beta}$ (iv) $E_3(F1, 0, G1)$ is a locally asymptotic stable point for formal firms. As usual, we consider the condition: $\sigma + \gamma G_1 > \alpha \overline{F}_1$ and

$$(a + \overline{F}_1)^2 \{\mu K + (2r + d)\overline{F}_1\} + G_1 K\{a(\beta + 2a\delta) + 2\delta \overline{F}_1(2a + \overline{F}_1)\}$$
$$> (a + \overline{F}_1)K\{(d + r)(a + \overline{F}_1) + l\beta \overline{F}_1\}$$

(v) The government free equilibrium $E_4(\overline{F}_2, F_2, 0)$ is locally asymptotically stable if $\frac{\alpha r \pi K}{(2r\pi + \alpha K)} < \sigma < \alpha K$ and $F_2 < \frac{F_{2a}}{F_{2b}}$, where

$F_{2a} = a\alpha\{d\sigma + (\mu - d)\alpha K\} + \sigma\{\alpha K\mu + d\sigma - \alpha K(d + l\beta)\}$ and

$F_{2b} = \alpha(a\alpha + \sigma)\{(m - n)\gamma K - d\pi\}$.

The theorem above explains the nature and behavior of the local equilibria system points through a local stability analysis. Analyzing the coexistence point $E^*$ is much more interesting than all the derived point s. The point explains the interaction and interlink between the three selected agents within the economy. It gives more information about the government's position when it imposes taxes on both formal and informal firms simultaneously and in the same economy.

### 3.3.4. Global stability

The equilibrium $E_2$ is the global asymptotic stable point with more economic effects. We provide some sufficient essential conditions for the point before we provide a detailed exploration. Point $E_2$ is firm-free, implying no firms to impose a tax on. This means that the government will survive through sources other than corporate tax. At the same time, the firms will be enjoying their 100% profits without suffering from such impositions. We provide a theorem to better explain the global point analysis as in local stability.

**Theorem 3.4.1**

For $Y_{min} \geq 0$, where $Y_{min} = \inf_G Y(G)$ and

$F(G) = (G - G_0)^2 - \frac{1}{\delta K}\left(rKC_{11}\mu_{21} + dG_0(\mu_{21} + \pi\mu_{22})\right)$, the system will be globally asymptotic stable at a "firm-free equilibrium point ."We will provide the proof subsequently.

**Proof:** We constructed and applied the Lyapunov function as it is convenient and efficient for analyzing the global stability of the system around $E_2$,

$V_2(\overline{F_p}, F_p, G_t) = A_1\overline{F_.} + B_1 F + C_1 \int_{G_0}^{G} \frac{G-G_0}{G} dG$,,,,,,,,,,,,,(16), where $A_1, B_1, C_1$ are positive constants. Further, we have

$\frac{dV_2}{dt} = -C_1\delta(G - G_0)^2 + C_1\left(\frac{l\beta\overline{F}}{a+\overline{F}} + (m-n)\gamma F - \frac{d}{K}(\overline{F} + \pi F)\right) \times (G - G_0) + A_1\left[r\overline{F}\left(1 - \frac{\overline{F}+\pi I}{K}\right) - \alpha\overline{F}F - \frac{\beta\overline{F}G}{a+\overline{F}}\right] + B_1(\alpha\overline{F}F - \gamma FG - \sigma F) \ldots\ldots\ldots\ldots\ldots (17).$

Further calculus takes us to;

$$\frac{dV_2}{dt} \leq \alpha(B_1 - A_1)SI + \gamma((m-n)C_1 - B_1)IP + (C_1 l - A_1)\frac{\beta SP}{a+S}$$
$$- \frac{\overline{F} + \eta I}{K}(C_1 d(G_0 - G) - A_1 rS) + ArS - B\sigma$$
$$- C_1\delta(G - G_0)^2 \ldots\ldots\ldots\ldots (18)$$

For simplicity we let $A_1 = B_1 = 1$ and $C_1 = min\left\{\frac{1}{m-n}, \frac{1}{l}\right\} = \frac{1}{C_{11}}$, for all $m > n$. Using these conditions, we thus get the following;

$$\frac{dV_2}{dt} \leq \frac{\bar{F} + \pi F}{K}\left(\frac{d(G_0 - G)}{C_{11}} - r\bar{F}\right) + (r\bar{F} - \sigma F) - \frac{\delta}{C_{11}}(G - G_0)^2 \ldots \ldots \ldots (19).$$

From equation (18) derived above, we managed to show that $\frac{dV_2}{dt} \leq 0$ if

$$F(G) = (G - G_0)^2 - \frac{1}{\delta K}(rKC_{11}\mu_{21} + dG_0(\mu_{21} + \pi\mu_{22})) \geq 0, \ldots \ldots (20)$$

Finally, we get; $F_{min} \geq 0$, and the proof is over. The mathematics of the tax-revenue system can better be explained by the following theorem, whose proof is not included in this paper.

**Theorem 3.4.2**

The system in (9) needs to be asymptotically stable at a global scale, and the following are sufficient conditions

(1) $H_1(0), H_2(0)$ are non-negative and

(2) $K\gamma n > d\pi + K\gamma n$

where

$$H_1(\bar{F}) = \frac{r}{K} + \frac{C_*d}{(2K)} - \frac{\beta(2P_*b + alC_*)}{2(a + \bar{F})(a + \bar{F}_*)}$$

$$H_2(\bar{F}) = \delta + \frac{C_*d}{2K} - \frac{a\beta lC_*}{2(a + \bar{F})(a + \bar{F}_*)}$$

and $C_* = \frac{K\gamma}{\alpha(K\gamma(m-n) - d\pi)}$.

### 3.3.5. Comments on the equilibrium points

This section provides the contextual meaning of each critical point derived above.

Our first critical point is the trivial one, $E_0(0, 0, 0)$. The point is impossible and infeasible. This is because, in both the short and long run, an economy can only exist with both firms (formal and informal) and the government (in terms of their revenue contributions). This means that it is not possible to have a null-component tax-revenue system. The second point is $E_1(K, 0, 0)$. This denotes economic stability, and it is feasible primarily in developed economies. The tax-revenue system will beat its total capacity. There are no revenue shortages; in a more precise sense, the government is in a bumper harvest. Thirdly, we have $E_2(0, 0, G)$. It is a case where the whole system consists of only the government. This may be because other firms have shut down, but it is also almost impossible. The point depicts that the government may be getting nothing from the firms due to tax hiding and economic slump effects like economic shutdown. When following critical events like war, global pandemics like COVID-19, or political instability, there may be severe/short or long-term shutdown periods.

We also have $E_3(\overline{F}, 0, G)$, which signifies a system with only two components; the government and the informal firms. This means that the government will only get its tax revenue from taxable informal businesses. This alternates to $E_4(0, F, G)$, a point with only the government and formal businesses. The informal companies may be hiding from the tax bracket, or they may all be unregistered, thus no tax payments. Another impossible point closely similar to our first trivial point is $E_5(\overline{F}, F, 0)$ which consists of only firms. No economy operates without the government; neither do they take a vacation nor shut down. Whether the economy is public, free, or mixed, the government $G$ component exists in any economy as they are the key regulators. Our key critical point is the coexistence point which consists of all the components and is denoted as; $E_6^*(\overline{F}, F, G)$. The point shows that the government can get its tax revenue from both formal and informal firms simultaneously. It is known as a complete system, unlike other points which depict an incomplete system. It is a feasible and fiscally efficient point, and thus, it should be maintained. Special attention, measures, and policies should be put in place to keep up the indefinite existence of the point. The point is considered a healthy economic point and should be optimized, as explained in the subsequent section.

4. **Application of optimal tax-revenue control system**

Our above-modeled system consists of the government and informal and formal firms. The aim is to model how the government can harvest Revenue from taxable firms. It is economically accurate that the government aims to maximize its revenue sums and proceeds with its harvest from the firms. However, the government is facing to some extent challenges of firms (informal) hiding from the tax levies. Such cases are expected in volatile economies like South Africa, where firms will try to survive. The operation and availability of unregistered firms also evidence this. The government must optimize the tax-revenue system so as to get as much output as possible. The government can best optimize its tax-revenue output by controlling its formal and informal firms to adhere to tax payments. The system (9) is restructured using the assumption that it is letting u be the penalty fee charged to the firms who lately fail to pay the levied taxes regularly or hide from the tax bracket. The rate/ amount of the penalty charged to both the suspected informal and formal firms is increased by respective factors $2u\bar{F}$ and $1uF$. A further assumption is made on the penalty fee. The hypothesis states that penalty fees for legal firms are more potent than the ones for their informal counterparts.

Further, the penalty fees generally affect, more or less, the third parties (spillover), like consumers and the government itself. We assume that due to the penalty fee charge controls, $(u)$, the rate of tax revenue of the government grows at a rate $\epsilon_3$. We aim to maximize the government's tax-revenue harvests by ensuring that every firm should be registered and keep all its tax payments to date. We should maximize the square of the applying penalty to maximize the possible tax revenues while minimizing the side effects like firm shutdowns. An optimal level of penalty should be charged to mitigate the effects mentioned above. Thus optimization, see below.

Let $L(\bar{F}, F, G, u)$ be the Lagrangian function defined in terms of the government $(G)$, informal firms $(\bar{F})$, and formal firms $(F)$, which needs to be optimized such that.

$$L(\bar{F}, F, G, u) = v_1 \bar{F} + v_2 F + v_3 u^2 \ldots \ldots \ldots \ldots \ldots \ldots (21)$$

We then employ the Hamiltonian of the problem to maximize the revenue sums collected from the firms as follows;

$$H_f = L + \Psi_1 \frac{d\bar{F}}{dt} + \Psi_2 \frac{dF}{dt} + \Psi_3 \frac{dG}{dt} \ldots \ldots \ldots \ldots \ldots (22)$$

Where $\Psi_i(t)$ for $i = 1, 2, 3$ are known adjoint variables or the costate variables, which can be determined by solving the following system of differential equations:

$$\widehat{\Psi_1} = -\frac{\partial H_f}{\partial \bar{F}} = -v_1 + \left[\alpha F - r\left(1 - \frac{2S + \pi F}{K}\right) + \frac{\alpha\beta G}{(a+\bar{F})^2} + \epsilon_1 u\right]\Psi 1 - \alpha F \Psi_2$$

$$+ \left(\frac{dG}{K} - \frac{la\beta G}{(a+\bar{F})^2}\right)\Psi_3 \widehat{\Psi_2} = -\frac{\partial H_f}{\partial F}$$

$$= -v_2 + \alpha + \frac{r\pi}{K}\bar{F}\Psi_1 + (\gamma G - \alpha\bar{F} + \sigma + \epsilon_2 u)\Psi_2$$

$$+ \left(\frac{d\pi}{K} - (m-n)\gamma\right)G\Psi_3,$$

(23)

$$\widehat{\Psi_3} = -\frac{\partial H_f}{\partial G} = \frac{\beta \bar{F}}{a+\bar{F}}\Psi_1 + \gamma F \Psi_2 + \left[\mu + 2\delta G + \epsilon_3 u - \frac{l\beta \bar{F}}{a+\bar{F}} - (m-n)\gamma F - d\left(1 - \frac{\bar{F}+\pi F}{K}\right)\right]\Psi_3,$$ satisfying the conditions; $\Psi_i(t_1) = 0, i = 1,2,3$.

We assume that $\hat{\bar{F}}$, $\widehat{F, and\ G}$ are the optimum value of $\bar{F}, F, and\ G$, respectively. Also, let $\widehat{\Psi_1}$, $\widehat{\Psi 2, \Psi_3}$ be the solutions of the system (23).

We then concluded our optimality problem with a theorem below, and we provided its proof after that

**Theorem 5.1** Let $u * (t)$ be an optimal control for $t \in [0, t_1]$ such that

$J(\bar{F}(t), F(t), u * (t)) = \min_u[J(\bar{F}(t), F(t), u(t)]$ subject to the system of modified differential equations(24) that is,

$$\begin{cases} \frac{d\overline{F_p}(t)}{dt} = r\overline{F_p}(t)\left(1 - \frac{\overline{F_p(t) + \gamma F_p(t)}}{K}\right) - \alpha\overline{F_p(t)}F_p - \frac{\beta\overline{F_p}G(t)}{a+\overline{F_p}} - \epsilon_1 u\overline{F_p}(t) \\ \frac{d\overline{F_p}(t)}{dt} = \alpha\overline{F_p(t)}F_p - \gamma F_p G(t) - \sigma F_p - \epsilon_2 u F_p \\ \frac{dG_r}{dt} = \frac{\beta\overline{F_p F_p}G(t)}{a+\overline{F_p}} + (m-n)\gamma F G_r(t) + d_r\left(1 - \frac{\overline{F_p(t) + \gamma F_p(t)}}{K}\right)G_r(t) - \mu G_r(t) - \\ \sigma G_r^2 - \epsilon_3 u G_r \end{cases}$$ (24)

**Proof**: refer to Berestycki and Zilio's (2019) paper. Overall, we noted that the optimal control is bounded, and the maximization of the tax-revenue harvesting system is possible.

## 5. Conclusions

Our primary drive is to identify the interlink between the government and formal and non-formal corporate firms in taxation terms. Among several tax heads, corporate tax is one of the primary sources of Revenue for governments. As such, it is worth studying how it is harvested from the firms by the government and the after-effects it imposes on the firms. Formal and informal firms will eventually die out or default despite the size of the tax rate, especially under volatile and slump economic conditions. Under boom or recovery environments, firms might survive, but if their profits are well managed and closely looked at, they will suffer and cease to enjoy growth. The critical problem arising from tax efforts applied by the government is the long-term failure of firms and retarded contributions to economic growth and sub-themes like employment creation. From the analysis, we propose that the governments introduce a healthy tax rate to charge firms (both formal and informal) to ensure the long-lasting survival of these and the related spillover effects. Optimal tax rates can be set through the Lagrangian approach to optimization, which this study discussed above.